\documentclass{article}
\usepackage{amssymb}

\newtheorem{theorem}{Theorem}

\newtheorem{lemma}[theorem]{Lemma}

\newtheorem{proposition}[theorem]{Proposition}

\input{tcilatex}
\begin{document}

\title{Expanding Multi-Market Monopoly and Nonconcavity in the Value of
Information }
\author{Stefan Behringer\thanks{%
SciencesPo, Paris, France. E-mail stefan.behringer@sciencepo.fr.} \\
}
\maketitle

\begin{abstract}
I investigate a Bayesian inverse problem in the specific setting of a price
setting monopolist facing a randomly growing demand in multiple possibly
interconnected markets. Investigating the Value of Information of a signal
to the monopolist in a fully dynamic discrete model employing the
Kalman-Bucy-Stratonovich filter, we find that it may be non-monotonic in the
variance of the signal. In the classical static settings of the Value of
Information literature this relationship may be convex or concave, but is
always monotonic. The existence of the non-monotonicity depends critically
on the exogenous growth rate of the system.

\medskip 


\medskip 

\end{abstract}

\section{Introduction}

\bigskip

The issue of how the explicit introduction of information impacts
traditional economic analysis of, for example, competitive equilibrium or
monopolistic behaviour has been investigated fruitfully within the paradigm
of asymmetric information. The success of explaining non-perfectly
competitive outcomes has however led to a neglect of the issue of how
information itself can be considered valuable. A single \emph{Value of
Information}\ (VoI)\ literature does not exits and the issue spans over
diverse economic fields and even other disciplines that aim to discriminate
signal from noise. Of course any investigation of a Value of Information
needs a specific reference frame in which such a "value" may occur.\footnote{%
I am grateful to Eduardo Perez, Michel DeLara, Tilman B\"{o}rgers, Godfrey
Keller, Olivier Gossner, Junjie Zhou, Christian Ewerhart, Jose
Herrera-Velasquez and participants of Oligo21 and EURO2021 for comments.}

\ 

In economics, Wilson (1975) investigates a monopoly context in which \emph{%
information may be acquired} in a quadratic-normal problem leading to simple
objective function containing only the first two modes of the uncertainty.
Wilson finds that an optimal choice of the number of signals to be acquired
at some fixed cost (while complicated to solve for explicitly)\ can be shown
to lead to corner-solutions, so either no, or an infinite amount of signals
are optimally purchased. This creates a problem for the existence of
competitive equilibrium. The mirror image of this result implies that the
objective function is \emph{strictly monotonic} \ (convex)\ in the \emph{%
precision\ of the signals}. 

\bigskip

A literature which has also employed the monopoly setting to investigate the
generation of information is the early \emph{experimentation} literature
e.g. Mirman et. al. (1993). The focus in this literature is on the
possibility to influence the generation of information itself in order to
improve the firm's objective rather than the optimal use of signals as such.
Clearly a Value of Information that is not concave in the firm's action may
again cause corner-solutions leading to extreme results regarding optimal
experimentation.

\bigskip

An influential paper that has argued that the Value of Information may
generically exhibit nonconcavities is Radner \& \ Stiglitz (1984), albeit
for finite signal realizations only. This assumption precludes many
interesting applications that are subject to continuous signals.

\bigskip

Many other academic disciplines such as physics, mathematics, or engineering
care about information and have developed sophisticated techniques in areas
such as \emph{machine learning}, c.f. Cover \& \ Thomas (2006). Here however
the concept of \emph{value} often differs from the economists focus and,
most frequently, such considerations are simply absent.

\bigskip

Whether one finds non-concavities in the Value of Information clearly also
depends on the specification of objective function of the agent(s). Even if
results such as Radner \& \ Stiglitz point into the other direction, as
noted in Chade \& \ Schlee (2002), p.423 the common intuition remains:

\bigskip

\begin{quote}
"Why should information exhibit increasing marginal returns over some range?
While information as a commodity is admittedly special, it seems that there
ought to exist a rich class of problems for which the value of information
is concave."
\end{quote}

\bigskip

In this paper, as in Behringer (2021) we will adopt the classical
microeconomic focus of Mirman et. al. (1993), recently also investigated in
Weber and Nguyen (2018)\ and Weber (2019) in an information acquisition
context and concentrate on a single-agent monopoly setting. This framework
allows to transfer the issue of information and signals into a
value-to-the-firm context smoothly.

\bigskip

We find analytical conditions for which the Value of Information from
signals in a fully dynamic setting with a growing market and continuous
signals exhibits non-monotonicities in their precision, contrary to the
finding of Wilson (1975), thereby adding some scepticism about the above
intuition. This implies that there exists a range of parameters in our
dynamic setting for which less precise information is beneficial for the
agent. Information is updated using the Kalman-Bucy-Stratonovich filter as
in Slade's (1989) Bertrand dupoly setting.

\bigskip

A paper that has previously noticed that less precise information may be
beneficial for a firm is Gal-Or (1988) which contrasts with Vives (1984) in
competitive settings, which have previously been investigated in Ponssard
(1976). However in Gal-Or the mechanism by which less precise information
about costs may increase firm's profits results from the competitive
situation only and the effect of information on the incentives for \emph{%
information sharing in oligopoly } that will allow for a plethora of
outcomes depending on the specification of the uncertainty, see e.g. Raith
(1996).

\bigskip

\section{The model}

The monopoly game has the following timing: The firm has initial priors over
an unknown state variable given by $(\mu _{0},\sigma _{0}^{2})$ about the
true market potential $\theta _{t}$ which has a known evolution over time.

\bigskip

In each period $t$ the monopolist sets a price $p_{t}$ according to its
present belief $\mu _{t}$ about the market potential which generates an
undiscounted noisy per-period profit $\Pi _{t}.$ The monopolist then
observes its own demand $q(\theta _{t},\mu _{t},\cdot )$ which acts as a
noisy signal about the true $\theta _{t}.$\ This signal can be used to \emph{%
learn} about the market potential and find the updated prior moments $(\mu
_{t+1},\sigma _{t+1}^{2})$ for the optimal prices in the next period.%
\footnote{%
We assume that firms are \textit{myopic} in the sense of Aghion et. al.
(1988) so that they ignore that by manipulating prices today they can gain
information about demand tomorrow but can use past sale data to update their
prices which may alternatively be viewed as a sequence of firms. Non-myopic
firms are treated in section 4.}

\bigskip

We thus have to solve for a \emph{dynamic pricing strategy} given a changing
market potential. The state variable \emph{for this single market potential}
changes over time as%
\[
\theta _{t+1}=d\theta _{t}+fe_{t+1} 
\]%
where $d,f>0$ are free parameters and the i.i.d. error $e_{t}$ is normalized
to $e_{t}\sim N(0,1)$ where $\sigma _{e}^{2}=1$ is w.l.o.g. as non-unity
variances can be embedded in $f$. To investigate true growth we further
assume that $d>1.$

\bigskip

At some time $t$ the monopolist operates in market $i=1,...,n$ choosing
prices against $n$ unknown demand functions, each of the linearized form%
\[
q_{i}=\theta _{i}-p_{i}+\gamma _{i} 
\]%
where the noises $\gamma _{i}\sim N(0,1)\,$\ are independent from $e_{t}$.

\bigskip

Operating at marginal cost $c,$ the monopolist will optimally choose a price 
\[
p_{i}=\frac{1}{2}\left( \theta _{i}+c+\gamma _{i}\right) 
\]%
in each market and obtain a per-period profit of%
\[
\Pi ^{\ast }=\frac{\left( \theta _{i}-c+\gamma _{i}\right) ^{2}}{4} 
\]%
Expectation of these profits in period $t$ absent learning would simply be 
\[
E\left \{ \Pi _{i}^{\ast }\right \} =\frac{\left( \mu _{i}-c\right) ^{2}}{4}%
. 
\]%
However the myopic monopolist knows the parameters of the model and can
observe sales in each period in each market.

\bigskip

As we allow for the market potential in one market to also change when the
potential in another market changes and to also depend on other markets
variations it is more convenient to employ \emph{vector notation}. The
market potentials will then change according to

\begin{equation}
\theta _{t+1}=D\theta _{t}+Fe_{t+1}  \label{G}
\end{equation}%
where now $D$ and $F$ are $n\times n$ matrices and $\theta _{t+1}$ and $%
e_{t} $ with $e_{t}\sim N(0,I_{n})\,$\ are $n\times 1$ vectors with $I_{n}$
denoting an $n\times n$ identity matrix. We assume the potential prior to
also be normally distributed $\theta _{0}\sim N(\mu _{0},\Sigma _{0}).$ All
matrices are invertible and transposes are indicated with a prime.

\bigskip

\begin{lemma}
\bigskip The conditional variance of the prior, given previous signals $%
S_{0},...S_{t-1}$ is%
\[
Var(\left. \theta _{t}\right \vert S_{0},...S_{t-1})=D\Sigma _{t}D^{\prime
}+FF^{\prime }. 
\]
\end{lemma}

\textbf{Proof:}

Standard, e.g., Veldkamp (2011).$\blacksquare $

\bigskip

\subsection{The learning process}

The learning process from observing sales allows for conventional Bayesian
updating with normal variables which in this dynamic setting becomes the 
\emph{Kalman-Bucy-Stratonovich filter} approach.

\bigskip

In a single isolated market in each period $t$ observed market demand
depends on the chosen price and therefore the belief about the market
potential as%
\[
q_{t}=\theta _{t}-\frac{\mu _{t}+c}{2}+\gamma _{t} 
\]%
which can be transformed into an \emph{unbiased signal} as%
\[
\left. \tilde{S}_{t}\right \vert \theta _{t}=q_{t}+\frac{\mu _{t}+c}{2}%
=\theta _{t}+\gamma _{t}. 
\]

\bigskip

Allowing for demand interrelations, e.g. competing products,
cannibalization, substitutive effects etc. by assuming that demands/signals
in one market can depend on other markets demands either systematically or
by idiosyncratic (error)\ realizations of both previous and contemporary
periods. This assumption extends the signal system to: 
\begin{equation}
S_{t}=G\theta _{t}+H\gamma _{t}  \label{S}
\end{equation}%
where the signal $S_{t}$ is an $n\times 1$ vector and $G$ and $H$ are $%
n\times n$ matrices with an $n\times 1$ error vector $\gamma _{t}\sim
N(0,I_{n})\bigskip .$

\begin{lemma}
\bigskip The variance of the unbiased signal is%
\[
Var(G^{-1}S_{t})=\Sigma _{t}+G^{-1}HH^{\prime }(G^{\prime })^{-1}. 
\]
\end{lemma}

\textbf{Proof:}

Premultiplying (\ref{S}) by $G^{-1}$ we have%
\begin{eqnarray*}
Var(G^{-1}S_{t}) &=&Var(\theta _{t})+Var(G^{-1}H\gamma _{t}) \\
&=&E\left \{ (\theta _{t}-\mu _{t})(\theta _{t}-\mu _{t})^{\prime }\right \}
+E\left \{ (G^{-1}H\gamma _{t})(G^{-1}H\gamma _{t})^{\prime }\right \} \\
&=&\Sigma _{t}+G^{-1}HH^{\prime }(G^{\prime })^{-1}
\end{eqnarray*}%
by independence and $\gamma _{t}\sim N(0,I_{n})\bigskip $ where $\mu _{t}$
is the belief about $\theta _{t}$ conditional on all previous signals, i.e.%
\[
\mu _{t}\equiv E\left \{ \left. \theta _{t}\right \vert
S_{0},...,S_{t-1}\right \} .\blacksquare 
\]

\begin{lemma}
\bigskip Optimal Bayesian beliefs about the unknown market potential can be
found from the system 
\[
\mu _{t+1}=K_{t}S_{t}+(D-K_{t}G)\mu _{t} 
\]%
where $K_{t}$ denotes the Kalman-Bucy-Stratonovich Filter (gain)\ which can
be determined as 
\[
K_{t}=D\Sigma _{t}G^{\prime }\left[ G\Sigma _{t}G^{\prime }+HH^{\prime }%
\right] ^{-1} 
\]%
with the conditional belief variance recursively determined as%
\[
\Sigma _{t+1}=D\Sigma _{t}D^{\prime }+FF^{\prime }-K_{t}G\Sigma
_{t}D^{\prime }. 
\]
\end{lemma}

\bigskip

\pagebreak

\textbf{Proof:}

Bayesian updating has a linear form due to the normality assumptions. Hence%
\[
\mu _{t+1}=DK_{t}^{^{\prime }}S_{t}+(D-DK_{t}^{^{\prime }}G)\mu _{t} 
\]%
and let's define the Kalman gain\ as $K_{t}=DK_{t}^{^{\prime }}$ so that it
includes the known market growth. Now 
\[
\mu _{t+1}=K_{t}S_{t}+(D-K_{t}G)\mu _{t} 
\]%
describes an optimal learning process. As%
\[
\Sigma _{t}=E\left \{ (\theta _{t}-\mu _{t})(\theta _{t}-\mu _{t})^{\prime
}\right \} 
\]%
and 
\[
\theta _{t+1}=D\theta _{t}+Fe_{t+1} 
\]%
we find the conditional variance of the prior in $t+1$ as%
\begin{eqnarray*}
\Sigma _{t+1} &=&E\left \{ \left. (\theta _{t+1}-\mu _{t+1})(\theta
_{t+1}-\mu _{t+1})^{\prime }\right \vert \theta _{0},...,\theta _{t}\right \}
\\
&=&E\left \{ \left. 
\begin{array}{c}
(Fe_{t+1}-K_{t}S_{t}-(D-K_{t}G)\mu _{t})\times \\ 
(Fe_{t+1}-K_{t}S_{t}-(D-K_{t}G)\mu _{t})^{\prime }%
\end{array}%
\right \vert \theta _{0},...,\theta _{t}\right \} \\
&=&(D-K_{t}G)\Sigma _{t}(D-K_{t}G)^{\prime }+FI_{n}F^{\prime
}+K_{t}HI_{n}H^{\prime }K_{t}^{\prime } \\
&=&D\Sigma _{t}D^{\prime }-K_{t}G\Sigma _{t}D^{\prime }-D\Sigma
_{t}(K_{t}G)^{\prime }+K_{t}G\Sigma _{t}(K_{t}G)^{\prime }+FF^{\prime
}+K_{t}HH^{\prime }K_{t}^{\prime } \\
&=&(D-K_{t}G)\Sigma _{t}D^{\prime }-D\Sigma _{t}G^{\prime }K_{t}^{\prime }+%
\underbrace{K_{t}(G\Sigma _{t}G^{\prime }+HH^{\prime })}_{D\Sigma
_{t}G^{\prime }}K_{t}^{\prime }+FF^{\prime } \\
&=&(D-K_{t}G)\Sigma _{t}D^{\prime }+FF^{\prime }
\end{eqnarray*}%
with the Kalman gain $K_{t}$ to be determined via the orthogonality
principle. Transforming

\begin{eqnarray*}
\Sigma _{t+1} &=&(D-K_{t}G)\Sigma _{t}D^{\prime }-D\Sigma _{t}G^{\prime
}K_{t}^{\prime }+K_{t}(G\Sigma _{t}G^{\prime }+HH^{\prime })K_{t}^{\prime
}+FF^{\prime } \\
&=&D\Sigma _{t}D^{\prime }-K_{t}G\Sigma _{t}D^{\prime }-D\Sigma
_{t}G^{\prime }K_{t}^{\prime }+K_{t}(G\Sigma _{t}G^{\prime }+HH^{\prime
})K_{t}^{\prime }+FF^{\prime }
\end{eqnarray*}%
Minimizing this variance, noting that the trace of a matrix is equal to the
trace of its transpose, thus%
\[
T[\Sigma _{t+1}]=T[D\Sigma _{t}D^{\prime }]-2T[K_{t}G\Sigma _{t}D^{\prime
}]+T[K_{t}(G\Sigma _{t}G^{\prime }+HH^{\prime })K_{t}^{\prime }] 
\]%
we differentiate w.r.t. $K_{t}$ and set to zero to find%
\[
\frac{dT[\Sigma _{t+1}]}{dK_{t}}=-2[G\Sigma _{t}D^{\prime }]^{\prime
}+2[K_{t}(G\Sigma _{t}G^{\prime }+HH^{\prime })]=0 
\]%
or%
\[
K_{t}=D\Sigma _{t}G^{\prime }\left[ G\Sigma _{t}G^{\prime }+HH^{\prime }%
\right] ^{-1} 
\]%
which leads to the belief updating signal weighting as%
\[
\mu _{t+1}=K_{t}GS_{t}+(D-K_{t}G)\mu _{t} 
\]%
and%
\[
\Sigma _{t+1}=D\Sigma _{t}D^{\prime }+FF^{\prime }-D\Sigma _{t}G^{^{\prime
}}(G\Sigma _{t}G^{\prime }+HH^{\prime })^{-1})G\Sigma _{t}D^{\prime
}.\blacksquare 
\]

\begin{lemma}
\bigskip The normalized Kalman gain $D^{-1}K_{t}$ represents the weight to
be put on the signal in Bayesian updating.
\end{lemma}

\textbf{Proof:}

Conditional on $\theta _{t\text{ }}$the signal has the variance of $%
G^{-1}HH^{\prime }(G^{\prime })^{-1}.$ The variance of the prior beliefs
about $\theta $ at time $t$ is $\Sigma _{t}$ so by linear/normal Bayesian
updating the agent will give a weight to the unbiased signal $G^{-1}S_{t}$
(with $G$ known) given by 
\[
\Sigma _{t}\left[ \Sigma _{t}+G^{-1}HH^{\prime }(G^{\prime })^{-1}\right]
^{-1} 
\]%
so the total weighted signal is%
\[
G^{-1}y_{t}\Sigma _{t}\left[ \Sigma _{t}+G^{-1}HH^{\prime }(G^{\prime })^{-1}%
\right] ^{-1}. 
\]%
The weight can be transformed by adding identities $(G^{\prime
})^{-1}G^{^{\prime }}=GG^{-1}=I$%
\begin{eqnarray*}
&&\Sigma _{t}(G^{\prime })^{-1}G^{^{\prime }}\left[ \Sigma
_{t}+G^{-1}HH^{\prime }(G^{\prime })^{-1}\right] ^{-1}GG^{-1} \\
&=&\Sigma _{t}G^{^{\prime }}\left[ G\Sigma _{t}G^{\prime }+HH^{\prime }%
\right] ^{-1}G
\end{eqnarray*}%
so the total weighted signal is%
\begin{eqnarray*}
&&G^{-1}S_{t}\Sigma _{t}G^{^{\prime }}\left[ G\Sigma _{t}G^{\prime
}+HH^{\prime }\right] ^{-1}G \\
&=&S_{t}\Sigma _{t}G^{^{\prime }}\left[ G\Sigma _{t}G^{\prime }+HH^{\prime }%
\right] ^{-1} \\
&=&S_{t}D^{-1}K_{t}
\end{eqnarray*}%
so that where $D^{-1}K_{t}$ represents the weight to be put on a signal in
period $t$.$\blacksquare $

\bigskip

\bigskip

We next focus on the expected profits of the monopolist:

\begin{lemma}
The firm obtains an expected profit in period $t$ of%
\[
E\left \{ \Pi _{t}\right \} =\frac{(D^{t}\mu _{t}-cI)^{2}}{4}+\frac{1}{4}%
K_{t-1}\Sigma _{t}K_{t-1}^{\prime } 
\]%
where the second term gives the pecuniary Value of Information of an
additional signal%
\begin{eqnarray*}
&&VoI_{t}=K_{t-1}\Sigma _{t}K_{t-1}^{\prime }= \\
&&D\Sigma _{t}G^{\prime }\left[ G\Sigma _{t}G^{\prime }+HH^{\prime }\right]
^{-1}\left[ (D-K_{t}G)\Sigma _{t}D^{\prime }+FF^{\prime }\right] \left[
D\Sigma _{t}G^{\prime }\left[ G\Sigma _{t}G^{\prime }+HH^{\prime }\right]
^{-1}\right] ^{\prime }.
\end{eqnarray*}
\end{lemma}

\bigskip

\textbf{Proof:}

In each market the monopolist has an expected per-period profit of 
\[
E\left \{ \Pi _{t}\right \} =\frac{1}{4}E\left \{ (D^{t}\mu
_{t}-cI)^{2}\right \} 
\]%
where again profit is an $n\times 1$ vector with $D^{t}$ denoting the $t$th
standard power of the $D$ matrix, i.e. $D^{t}=\Pi _{i=1}^{t}D_{i}$. As 
\begin{eqnarray*}
\frac{1}{4}E\left \{ (\mu _{t}-cI)^{2}\right \} &=&\frac{1}{4}%
Var(K_{t-1}GS_{t}+(D^{t}-K_{t-1}G)\mu _{t-1}) \\
&&+\frac{1}{4}\left[ E\left \{ K_{t-1}GS_{t}+(D^{t}-K_{t-1}G)\mu
_{t-1}-cI\right \} \right] ^{2} \\
&=&\frac{1}{4}Var(K_{t-1}GS_{t})+\frac{1}{4}\left[ D^{t}\mu _{t-1}-cI\right]
^{2} \\
&=&\frac{1}{4}\left[ D^{t}\mu _{t-1}-cI\right] ^{2}+\frac{1}{4}%
GK_{t-1}\Sigma _{t}K_{t-1}^{\prime }G^{\prime }.\blacksquare
\end{eqnarray*}

We normalize $G=I$ and focus on 
\[
\frac{1}{4}\left[ D^{t}\mu _{t-1}-cI\right] ^{2}+\frac{1}{4}K_{t-1}\Sigma
_{t}K_{t-1}^{\prime }. 
\]

Note that the simple Kalman filter setup cannot accommodate a multiplicative
error case as for example investigated in Behringer (2021). The problem is
that the estimated variance (that goes into the next period) will depend on
the particular realization of the signal (via the action of the monopolist)
and is no longer an unconditional error covariance as in the white noise\
case. One approach here would be to consider conditionally Gaussian models,
see for example Harvey, (1989) p.155ff, but this lends itself to asymptotic
econometric testings only.

\bigskip

In the dynamic setting investigated in this paper however we are still able
to derive a non-monotonicity in Value of Information even in the white noise
case.

\bigskip

\section{\protect \bigskip $VoI_{t}$ with increasing signal noise}

The variance of the prior is recursively given by

\[
\Sigma _{t+1}=D\Sigma _{t}D^{\prime }+FF^{\prime }-D\Sigma _{t}G^{^{\prime
}}(G\Sigma _{t}G^{\prime }+HH^{\prime })^{-1}G\Sigma _{t}D^{\prime } 
\]%
The system given in Lemma 3 can be guaranteed to possess asymptotically
stability, see Harvey, 1989, p.118ff.

\bigskip

In the following we focus on a single market and denote a generic element of
a matrix $X\, \ $as $X_{ii(t)}.$ Alternatively one may assume that $X$
denotes \emph{diagonal matrices} so that one could for example take roots by
simply taking the roots of the terms on the major diagonal. Also then the
elements of the updated variance $\Sigma _{t+1}$ are increasing and concave
in $\Sigma _{t}$ and will be converging asymptotically over time.

\bigskip

This \emph{long-run (steady-state) variance} in any market can be calculated
as%
\[
\bigskip \Sigma _{ii}^{\ast }=\frac{1}{2G_{ii}^{2}}\left( 
\begin{array}{c}
\sqrt{H_{ii}^{2}\left( H_{ii}^{2}\left( D_{ii}-I_{ii}\right) ^{2}\left(
D_{ii}+I_{ii}\right) ^{2}+\allowbreak 2F_{ii}^{2}G_{ii}^{2}\left(
D_{ii}^{2}+I_{ii}\right) \right) \allowbreak +F_{ii}^{4}G_{ii}^{4}} \\ 
+H_{ii}^{2}\left( D_{ii}-I_{ii}\right) \left( D_{ii}+I_{ii}\right)
+F_{ii}^{2}G_{ii}^{2}%
\end{array}%
\right) 
\]%
which is approximately \emph{quadratic} in the signal noise. In the
"no-growth" case $D=I$ with diagonal element $I_{ii}=1$ we find%
\[
\Sigma _{ii}^{\ast }\left \vert _{D=I}\right. =\frac{F_{ii}}{2G_{ii}}\sqrt{%
F_{ii}^{2}G_{ii}^{2}+4H_{ii}^{2}}+F_{ii}^{2}G_{ii}^{2} 
\]%
to be approximately \emph{linear} in the signal noise.

\bigskip

\subsection{Pro-rata case $VoI_{t}\left[ K_{t-1}^{\prime }\right] ^{-1}$}

As noted in Radner \& Stiglitz (1984) in linear prediction settings monotone
transformations of the information structure (e.g. by a square) may alter
findings. This motivates our investigation of a time $t$ pro-rata time $VoI$%
, where the Kalman gain matrix is not multiplied with itself. While a good
economic interpretation of this concept may be hard to come by we include it
for its important didactic purposes. We find that:

\bigskip

\begin{lemma}
$VoI_{t}\left[ K_{ii,t-1}^{\prime }\right] ^{-1}=K_{ii,t-1}\Sigma
_{ii}^{\ast }$ has a global minimum at the lowest signal noise $H_{ii}=0$.
\end{lemma}

\bigskip

\textbf{Proof:}

The result is not obvious as $K_{ii,t-1},$ has a global maximum at $H_{ii}=0$
and $\Sigma _{ii}^{\ast }$, is convex in the signal noise with a global
minimum at $H_{ii}=0$ so their product needs to be explicitly determined.
The first order condition is multiplicative in $H_{ii}$ and thus shows the
existence of an extreme value. Calculating its second order condition at
this extreme value we find that

\[
\left. \frac{\partial ^{2}(K_{ii,t-1}\Sigma _{ii}^{\ast })}{\partial
H_{ii}^{2}}\right \vert _{H_{ii}=0}=(F_{ii}G_{ii})^{12}D_{ii}\left(
2G_{ii}^{2}+I_{ii}\right) ^{4}\left( D_{ii}-I_{ii}\right) \left(
D_{ii}+I_{ii}\right) >0 
\]%
which holds for any matrices as $D>I$\ holds element-wise.$\blacksquare $

\bigskip

This finding implies that the long-run conditional prior variance always
increases faster in the signal noise than the Kalman gain\ decreases. We
next show that this is no longer the case once we look at the full stage VoI
with a squared Kalman gain as is implied in the chosen monopoly setting.

\bigskip

\subsection{Full case $VoI_{t}$}

The effect on the full stage Value of Information is fundamentally different
once the square of the Kalman gain is taken into account. It turns out that
in this case the increasing variance of the prior term only dominates the
total effect for large values of signal noise, not small ones. This leads to
a \emph{non-monotonicity} in the monopoly $VoI_{t}$, implying that changes
in the signal noise have an ambiguous effect on expected profits the exact
direction of which will depend on the growth rate of the market. This is
formally shown next:

\bigskip

\begin{proposition}
$VoI_{t}=K_{ii,t-1}\Sigma _{ii}K_{ii,t-1}^{\prime }$ has an (local)\ extreme
value at the lowest signal noise $H_{ii}=0$ and there exists a range of
market growths $D_{ii}$ such that this extreme value is a maximum implying a
non-monotonicity of the $VoI_{t}$ with respect to the signal precision.
\end{proposition}

\bigskip

\textbf{Proof:}

The first order condition $\partial (K_{ii,t-1}\Sigma
_{ii}K_{ii,t-1}^{\prime })/\partial H_{ii}$ is again multiplicative in $%
H_{ii}.$The second derivative of the $VoI_{t}$ at the extreme value can be
found as 
\[
\left. \frac{\partial ^{2}(K_{ii,t-1}\Sigma _{ii}K_{ii,t-1}^{\prime })}{%
\partial H_{ii}^{2}}\right \vert _{H_{ii}=0}=\frac{D_{ii}^{2}}{G_{ii}^{4}}%
\left( 2G_{ii}^{2}(D_{ii}-I_{ii})(D_{ii}+I_{ii})+D_{ii}^{2}-3\right) 
\]%
so that the extreme value is a maximum if the growth satisfies 
\[
I_{ii}\leq D_{ii}\leq \sqrt{8G_{ii}^{2}+4G_{ii}^{4}+3}\left[
2G_{ii}^{2}+I_{ii}\right] ^{-1}. 
\]%
As $VoI_{t}$ is unbounded above with $D>I$ again holding element-wise, this
implies a non-monotonicity in signal noise which completes the proof. Note
that this result does not depend on the prior noise $F_{ii}$.$\blacksquare $

\bigskip

We have thus shown that even in a setup without a multiplicative signal
error as investigated in Behringer (2021), in a dynamic model with a growing
market we can find that the VoI to a monopoly firm may be non-monotonic in
the signal precision. In the former model without growth the conditional
prior variance is concave in the signal noise. In contrast in the present
dynamic setting with growth, the long run conditional prior variance is
convex in the signal noise.

\bigskip

In static settings, e.g. Wilson (1975) the weight to be put on the signal
will be strictly decreasing in the signal noise while the variance of the
prior is fixed and given. Hence the VoI in the static case in unambiguously
decreasing in this noise. In a dynamic setting with growth the priors will
be updated and the VoI at each stage will be subject to these updates. The
VoI is composed of the updated prior variance and the optimal weights. As in
the static setting the optimal weights will be decreasing in the signal
noise but now, contrary to static models, the updated belief prior will
increase in the signal noise.

\bigskip

\section{Non-myopic case}

\bigskip Let the random linear(ized) demand now be 
\[
q=\theta -B\beta p+H\gamma 
\]%
where $B$ and $H$ are known matrices, \emph{additive} noise is $\gamma \sim
N(0,I)$ and \emph{multiplicative} noise is $\beta \sim N(I,I)$ as in the
two-stage game of Behringer (2021) where the interplay of the two noises
results in a VoI non-monotonicity$.$ Note that this implies a continuum of
possible demand curves with varying slopes and intercepts. The demand
intercept is: 
\[
\theta _{t+1}=D\theta _{t}+Fe_{t+1} 
\]%
For this $AR(1)$ process $D$ and $F$ are known matrices and $\theta _{t}$
and $e_{t}$ with $e_{t}\sim N(0,I)\,$\ are vectors and $\theta _{0}\sim
N(\mu _{0},\Sigma _{0})$. 
\[
\mu _{t}=E_{t}\left \{ \theta _{t}\right \} \text{ and }\Sigma
_{t}=E_{t}\left \{ (\theta _{t}-\mu _{t})^{2}\right \} 
\]%
Firms thus observe a signal system:%
\[
S_{t+1}=\theta _{t}-B\beta _{t}p_{t}+H\gamma _{t} 
\]%
where learning is more complicated as the random effects cannot be
disentangled. The conditional signal variance is then: 
\[
Var(\left. S\right \vert \theta ,p)=Bp_{t}p_{t}^{\prime }B^{\prime
}+HH^{\prime } 
\]%
Optimization for per-period profit \emph{without learning} yields the
necessary condition for expected optimal \emph{myopic} prices as%
\[
\frac{dE_{t}\left \{ \Pi _{t}\right \} }{dp_{t}}=E_{t}\left \{ \frac{d\Pi
_{t}}{dp_{t}}\right \} =\mu _{t}+Bc-2Bp_{t}=0. 
\]

\begin{lemma}
Optimal Bayesian beliefs about the unknown market potential can be found
from the system 
\[
\mu _{t+1}=D\left[ K_{t}S_{t+1}+(I-K_{t})\mu _{t}\right] 
\]%
where $K_{t}$ denotes the (now growth independently defined)
Kalman-Bucy-Stratonovich gain 
\[
K_{t}=\Sigma _{t}\left[ \Sigma _{t}+Bp_{t}p_{t}^{\prime }B^{\prime
}+HH^{\prime }\right] ^{-1} 
\]%
with the conditional belief variance recursively determined as%
\[
\Sigma _{t+1}=D(I-K_{t})\Sigma _{t}D^{\prime }+FF^{\prime }=D\Sigma
_{t}D^{\prime }+FF^{\prime }-DK_{t}\Sigma _{t}D^{\prime }. 
\]
\end{lemma}

\bigskip

Full dynamic optimization via the \emph{Bellman equation} with prior
estimate $\mu _{t}$ and variance $\Sigma _{t}$ as state variables:%
\[
V(\mu _{t},\Sigma _{t})=\max_{p_{t}}E_{t}\left[ \Pi _{t}+\delta V(\mu
_{t+1},\Sigma _{t+1})\right] 
\]%
and constant discount factor $\delta \in (0,1).$ Thus $V(\mu _{t},\Sigma
_{t})=$%
\begin{equation}
\max_{p_{t}}E_{t}\left \{ \Pi _{t}+\delta V(D(K_{t}S_{t+1}+(I-K_{t})\mu
_{t}),D(I-K_{t})\Sigma _{t}D^{\prime }+FF^{\prime })\right \}  \nonumber
\end{equation}%
where now $K_{t}$ (and thus also $\Sigma _{t+1}$) depends on the choice of $%
p_{t}.$

\bigskip

\begin{lemma}
\bigskip The Bellman equation can be rewritten as%
\begin{eqnarray*}
&&E_{t}\left \{ \frac{d\Pi _{t}}{dp_{t}}\right \} +\frac{\frac{\partial
(K_{t}\sqrt{\Sigma _{t}+HH^{\prime }+Bp_{t}p_{t}^{\prime }B^{\prime }})}{%
\partial p_{t}}}{\frac{\partial (K_{t}\sqrt{\Sigma _{t}+HH^{\prime
}+Bp_{t}p_{t}^{\prime }B^{\prime }})}{\partial \Sigma _{t}}}\frac{\partial V%
}{\partial \Sigma _{t}}-E_{t}\left \{ \frac{\partial V}{\partial \Sigma
_{t+1}}\right \} \times \\
&&\delta D\left( \frac{\frac{\partial (K_{t}\sqrt{\Sigma _{t}+HH^{\prime
}+Bp_{t}p_{t}^{\prime }B^{\prime }})}{\partial p_{t}}}{\frac{\partial (K_{t}%
\sqrt{\Sigma _{t}+HH^{\prime }+Bp_{t}p_{t}^{\prime }B^{\prime }})}{\partial
\Sigma _{t}}}\left( I-\frac{\partial (K_{t}\Sigma _{t})}{\partial \Sigma _{t}%
}\right) +\frac{\partial (K_{t}\Sigma _{t})}{\partial p_{t}}\right)
D^{\prime }=0.
\end{eqnarray*}
\end{lemma}

\bigskip

\textbf{Proof:}

As the new signal $S_{t+1}$ is normal with mean $\mu _{t}$~and variance $%
\Sigma _{t}+HH^{\prime }+Bp_{t}p_{t}^{\prime }B^{\prime }$ we have that%
\[
\varepsilon _{t+1}=\frac{S_{t+1}-\mu _{t}}{\sqrt{\Sigma _{t}+HH^{\prime
}+Bp_{t}p_{t}^{\prime }B^{\prime }}} 
\]%
is standard normally distributed. Replacing $S_{t+1}$ in the \emph{Bellman
equation} yields 
\[
V(\mu _{t},\Sigma _{t})=\max_{p_{t}}E_{t}\left \{ 
\begin{array}{c}
\Pi _{t}+\delta V(D(\mu _{t}+K_{t}(\sqrt{\Sigma _{t}+HH^{\prime
}+Bp_{t}p_{t}^{\prime }B^{\prime }})\varepsilon _{t+1}), \\ 
D(I-K_{t})\Sigma _{t}D^{\prime }+FF^{\prime })%
\end{array}%
\right \} . 
\]%
The \emph{Euler condition} for this equation (with variables at $t$ known)
is:%
\begin{eqnarray*}
&&E_{t}\left \{ \frac{d\Pi _{t}}{dp_{t}}\right \} +\delta D\frac{\partial
(K_{t}(\sqrt{\Sigma _{t}+HH^{\prime }+Bp_{t}p_{t}^{\prime }B^{\prime }})}{%
\partial p_{t}}E_{t}\left \{ \frac{\partial V}{\partial \mu _{t+1}}%
\varepsilon _{t+1}\right \} - \\
&&\delta \frac{\partial (DK_{t}\Sigma _{t}D^{\prime })}{\partial p_{t}}%
E_{t}\left \{ \frac{\partial V}{\partial \Sigma _{t+1}}\right \} =0.
\end{eqnarray*}%
An application of the \emph{envelope theorem} to the Bellman equation%
\[
\max_{p_{t}}E_{t}\left \{ 
\begin{array}{c}
\Pi _{t}+\delta V(D(\mu _{t}+K_{t}(\sqrt{\Sigma _{t}+HH^{\prime
}+Bp_{t}p_{t}^{\prime }B^{\prime }})\varepsilon _{t+1}), \\ 
D(I-K_{t})\Sigma _{t}D^{\prime }+FF^{\prime })%
\end{array}%
\right \} 
\]%
yields%
\begin{eqnarray*}
\frac{\partial V}{\partial \Sigma _{t}} &=&\delta D\frac{\partial (K_{t}%
\sqrt{\Sigma _{t}+HH^{\prime }+Bp_{t}p_{t}^{\prime }B^{\prime }})}{\partial
\Sigma _{t}}E_{t}\left \{ \frac{\partial V}{\partial \mu _{t+1}}\varepsilon
_{t+1}\right \} + \\
&&\delta D\left( I-\frac{\partial (K_{t}\Sigma _{t})}{\partial \Sigma _{t}}%
\right) D^{\prime }E_{t}\left \{ \frac{\partial V}{\partial \Sigma _{t+1}}%
\right \} .
\end{eqnarray*}%
Rearranging and substituting this into the Bellman equation we find the
above expression.$\blacksquare $

\bigskip

\begin{lemma}
\bigskip The Euler equation can be simplified to%
\[
E_{t}\left \{ \frac{d\Pi _{t}}{dp_{t}}\right \} -\frac{2Bp_{t}B^{\prime
}\Sigma _{t}}{\Sigma _{t}+2Bp_{t}p_{t}^{\prime }B^{\prime }+2HH^{\prime }}%
\left( \frac{\partial V}{\partial \Sigma _{t}}-\delta DD^{\prime }E_{t}\left
\{ \frac{\partial V}{\partial \Sigma _{t+1}}\right \} \right) =0. 
\]
\end{lemma}

\bigskip

\textbf{Proof:}

\bigskip As optimal learning implies that 
\[
K_{t}=\Sigma _{t}\left[ \Sigma _{t}+Bp_{t}p_{t}^{\prime }B^{\prime
}+HH^{\prime }\right] ^{-1} 
\]%
calculating the derivatives we find that:%
\begin{eqnarray*}
&&\left( \frac{\frac{\partial (K_{t}(\sqrt{\Sigma _{t}+HH^{\prime
}+Bp_{t}p_{t}^{\prime }B^{\prime }})}{\partial p_{t}}}{\frac{\partial (K_{t}%
\sqrt{\Sigma _{t}+HH^{\prime }+Bp_{t}p_{t}^{\prime }B^{\prime }})}{\partial
\Sigma _{t}}}\left( I-\frac{\partial (K_{t}\Sigma _{t})}{\partial \Sigma _{t}%
}\right) +\frac{\partial (K_{t}\Sigma _{t})}{\partial p_{t}}\right) = \\
&&\frac{\frac{\partial (K_{t}(\sqrt{\Sigma _{t}+HH^{\prime
}+Bp_{t}p_{t}^{\prime }B^{\prime }})}{\partial p_{t}}}{\frac{\partial (K_{t}%
\sqrt{\Sigma _{t}+HH^{\prime }+Bp_{t}p_{t}^{\prime }B^{\prime }})}{\partial
\Sigma _{t}}}=-\frac{2Bp_{t}B^{\prime }\Sigma _{t}}{\Sigma
_{t}+2Bp_{t}p_{t}^{\prime }B^{\prime }+2HH^{\prime }}
\end{eqnarray*}%
so that the Euler equations simplifies.$\blacksquare $

\bigskip

The equation represents the trade-off of present (with myopically optimal
price)\ and future gains from optimal non-myopic pricing. As $\frac{\partial
V}{\partial \Sigma _{t}}$ and $\frac{\partial V}{\partial \Sigma _{t+1}}$
still depend on it, an explicit calculation of the optimal non-myopic price
is prevented.

\bigskip

We next proceed to show that an interplay between the multiplicative and the
additive demand error remains if the latter is large, as in the
non-monotonicity finding of Behringer (2021).

\bigskip

\begin{proposition}
With large signal noise $H_{ii}$ the optimal non-myopic price will still
depand on the degree of multiplicative demand variance indicating a
non-monotonicity in the VoI.
\end{proposition}

\bigskip

\textbf{Proof:}

Again the Ricatti recursion for $\Sigma _{t+1}$ converges to a \emph{%
long-run (steady-state) belief variance} it can be solved directly as%
\[
\Sigma _{ii}^{\ast }=\frac{1}{2}\left( 
\begin{array}{c}
\sqrt{%
\begin{array}{c}
\left( H_{ii}^{2}+B_{ii}^{2}p_{ii}^{2}\right) ^{2}\left(
D_{ii}-I_{ii}\right) ^{2}\left( D_{ii}+I_{ii}\right) ^{2} \\ 
+2F_{ii}^{2}(H_{ii}^{2}+B_{ii}^{2}p_{ii}^{2})\left( D_{ii}^{2}+I_{ii}\right)
+F_{ii}^{4}%
\end{array}%
} \\ 
\left( H_{ii}^{2}+B_{ii}^{2}p_{ii}^{2}\right) \left( D_{ii}-I_{ii}\right)
\left( D_{ii}+I_{ii}\right) +F_{ii}^{2}%
\end{array}%
\right) . 
\]%
The critical term in the Euler equation attains a limit for large signal
noise:%
\[
\lim_{H_{ii}\rightarrow \infty }\left( -\frac{2B_{ii}^{2}p_{t}\Sigma
_{ii}^{\ast }}{\Sigma _{ii}^{\ast }+2B_{ii}^{2}p_{ii}^{2}+2H_{ii}^{2}}%
\right) =-\frac{2B_{ii}^{2}p_{t}\left( D_{ii}-I_{ii}\right) \left(
D_{ii}+I_{ii}\right) }{D_{ii}^{2}+I_{ii}} 
\]%
so that while falling and convex in the growth rate for $D_{ii}>0,$ even for
a very unreliable signal the term still depends on the multiplicative noise
in the demand function implying an effect on \emph{any} optimal non-myopic
price as $E_{t}\left \{ d\Pi _{t}/dp_{t}\right \} $ is affine.$\blacksquare $

\bigskip

\section{Conclusion}

\bigskip

The two-stage analysis with multiplicative error investigated in Behringer
(2021) has found a non-monotonicity in the Value of information for the
monopolist resulting from the interplay of the variance of the
multiplicative error and the level of the prior for the market potential
demand and cost. A higher variance of the total signal noise (additive and
multiplicative)\ will eventually decrease the total Value of Information in
the two stage setup, in which there is only one signal and the conditional
variance is only updated once, to zero. This results from the signal weight
in the Bayesian process being strictly decreasing in the signal noise and
the posterior conditional variance in the second stage being increasing but
bounded.

\bigskip

In a fully dynamic framework where the moments of the estimates are state
variables, a non-monotonicity of the Value of information can still be found
when allowing for myopic agents and additive errors only. The long run
updated conditional variance of the prior resulting from a Bayesian learning
process will be increasing in the signal noise if market demand increases
over time. Despite the fact that the signal weight in the Bayesian process
resulting from the Kalman-Bucy-Stratonovich filter is strictly decreasing in
the signal noise the pro-rata stage VoI will strictly increase. However,
when squaring the weight in the full VoI analysis the decreasing weight will
dominate for low values of market growth and signal noise leading to a
non-monotonicity of the stage VoI in the signal noise, contrary to the
findings in static models such as Wilson (1975).

\bigskip

\pagebreak

With large signal noise an interplay of multiplicative and additive error
(which leads to a non-monotone VoI\ in Behringer 2021) even if
uncorrelated,\ prevails also in a fully dynamic, non-myopic setting. These
results imply that the concerns expressed in Radner \& \ Stiglitz (1984) are
robust to the introduction of continuous signals and true dynamics. The
overall intuition raised in Chade \& \ Schlee (2002) or Arrow (1985)
prevails but will have to be tested for even the classical quadratic
settings (here motivated by monopoly) which allow for a monetization of the
information concept. This may in some cases (including the present one)\
allow for analytical solutions and guidance.

\bigskip

\section{References:}

\bigskip

Aghion, P., Bolton, P., \& Jullien, B. (1988): "Learning through price
experimentation by a monopolist facing unknown demand", MIT WP, No. 491.

\bigskip

Arrow, K.J. (1985): "Information Structure of the Firm",\textit{\ American
Economic Review}, Vol .75, No. 2. pp. 303-307.

\bigskip

Behringer, S. (2021): "Multiplicative Normal Noise and Nonconcavity in the
Value of Information", \emph{Theoretical Economics Letters}, 11, pp. 116-124.

\bigskip

Chade, H. \& \ Schlee, E.E. (2002): "Another Look at the Radner-Stiglitz
Nonconcavity in the Value of Information", \textit{Journal of Economic Theory%
}, Vol. 107, pp. 421-452.

\bigskip

Cover, T.M. and Thomas, J.A. (2006): \textit{Elements of Information Theory}%
, 2nd, edition, Wiley-Interscience.

\bigskip

Gal-Or, E. (1988): "The Advantages of Imprecise Information", \textit{The
RAND Journal of Economics}, Vol. 19, No. 2, pp. 266-275.

\bigskip

Mirman, L.J., Samuelson, L. \& Urbano, A. (1993): "Monopoly
Experimentation", \emph{International Economics Review}, Vol. 34, No. 3, pp.
549-563.

\bigskip

Ponssard, J.-P. (1975): "On the concept of the value of information in
competitive situations" \emph{Management Sciences}, Vol. 22, pp. 739-748

\bigskip

Radner, R. and Stiglitz, J. (1984): "A nonconcavity in the value of
information", in \textit{Bayesian Models of Economic Theory} (Boyer, M. and
Kihlstrom ,R., Eds.), pp. 33--52, Elsevier Amsterdam.

\bigskip

Raith, M. (1996): "A General Model of Information Sharing in Oligopoly", 
\textit{Journal of Economic Theory}, Vol. 71, pp. 260-288.

\bigskip

Harvey, A.C. (1989): \textit{Forecasting, structural time series models and
the Kalman filter}, Cambridge University Press.

\bigskip

Slade, M. (1989): "Price Wars in Price-Setting Supergames", \textit{Economica%
}, Vol. 56, No. 223, pp. 295-310.

\bigskip

Veldkamp, L. (2011): \textit{Information Choice in Macroeconomics and Finance%
}, Princeton University Press.

\bigskip

Vives, X. (1984): "Duopoly, Information Equilibrium: Cournot and Bertrand", 
\textit{Journal of Economic Theory}, Vol. 34, pp. 71-94.

\bigskip

Weber, T. (2019): "Dynamic Learning in Markets: Pricing, Advertising, and
Information Acquisition", HICSS Conference Proceedings, p.6628-6637.

\bigskip

Weber, T. and Nguyen, V.A. (2018): "A linear-quadratic Gaussian approach to
dynamic information acquisition", \textit{European Journal of Operations
Research}, Vol 270, Issue 1, pp. 260--281.

\bigskip

Wilson, R. (1975): "Informational economies of scale", \textit{The Bell
Journal of Economics}, Vol. 6, No.1, pp.184-195.

\bigskip \bigskip

\end{document}